\documentclass[twocolumn,10pt,aps,prl,nobibnotes]{revtex4-1}
\usepackage{amssymb, amsmath, amsthm}
\usepackage{graphicx,import}
\usepackage{units}

\usepackage{upgreek}
\usepackage{color}
\usepackage[usenames,dvipsnames]{xcolor}
\usepackage{url}
\begin{document}
\title{Probing an Excited-State Atomic Transition Using Hyperfine Quantum Beat Spectroscopy}
\author{C. G. Wade}\email{c.g.wade@durham.ac.uk}
\author{N. \v{S}ibali\'c}
\author{J. Keaveney}
\author{C. S. Adams}
\author{K. J. Weatherill}
\affiliation{Joint Quantum Centre (JQC) Durham-Newcastle, Department of Physics, Durham University, South Road, Durham, DH1 3LE, United Kingdom}
\begin{abstract}
\noindent We describe a method to observe the dynamics of an excited-state transition in a room temperature atomic vapor using hyperfine quantum beats. Our experiment using cesium atoms consists of a pulsed excitation of the D$_2$ transition, and continuous-wave driving of an excited-state transition from the 6P$_{3/2}$ state to the 7S$_{1/2}$ state. We observe quantum beats in the fluorescence from the 6P$_{3/2}$ state which are modified by the driving of the excited-state transition. The Fourier spectrum of the beat signal yields evidence of Autler-Townes splitting of the 6P$_{3/2}$, $F$~=~5 hyperfine level and Rabi oscillations on the excited-state transition. A detailed model provides qualitative agreement with the data, giving insight to the physical processes involved.
\end{abstract}
\date{\today}
\maketitle


\section{Introduction}
\label{sec:intro}

\noindent Excited-state transitions in atomic systems are finding an increasing range of applications including quantum information~\cite{HarocheBook}, optical filters~\cite{ESFADOF}, electric field sensing~\cite{moha08,sedl12,holl14} and quantum optics~\cite{prit13,dudi12}. They are also used for state lifetime measurements~\cite{shen08}, frequency up-conversion \cite{meij06}, the search for new stable frequency references~\cite{bret93,abel09} and multi-photon laser cooling~\cite{wu09}. However excited-state transitions are inherently more difficult to probe than ground-state transitions, especially if the lower state is short-lived. It is possible to probe an excited-state transition directly if the dipole moment is large enough~\cite{tana12}, but more commonly excited-state character is observed by mapping onto ground-state transitions using electromagnetically induced transparency (EIT) in a ladder configuration. Using EIT it is possible to probe even relatively weak excited-state transitions, such as those to highly excited Rydberg states~\cite{moha07,weat08,carr12a}. Nanosecond timescales have also been probed, effectively `freezing out' the motion of thermal atoms~\cite{hube11}.

EIT involving Rydberg states has paved the way to recent advances in non-linear and quantum optics \cite{prit13} as the strong interactions among the Rydberg atoms lead to large optical non-linearities, even at the single-photon level \cite{dudi12,peyr12,maxw13}. In room temperature Rydberg gases, the atomic interactions can lead to a non-equilibrium phase transition \cite{carr13} and evidence for strong van der Waals interactions has been observed~\cite{balu13}. Despite the considerable successes of ladder EIT, there is a particular class of energy level schemes for which ladder EIT cannot be observed in a Doppler-broadened medium. Specifically, when the upper transition wavelength is longer than the lower (`inverted-wavelength' system)~\cite{Boon99,urvo13}, the transparency window is absent as it is smeared out by velocity averaging. 

\begin{figure}
\includegraphics[width=3.4in]{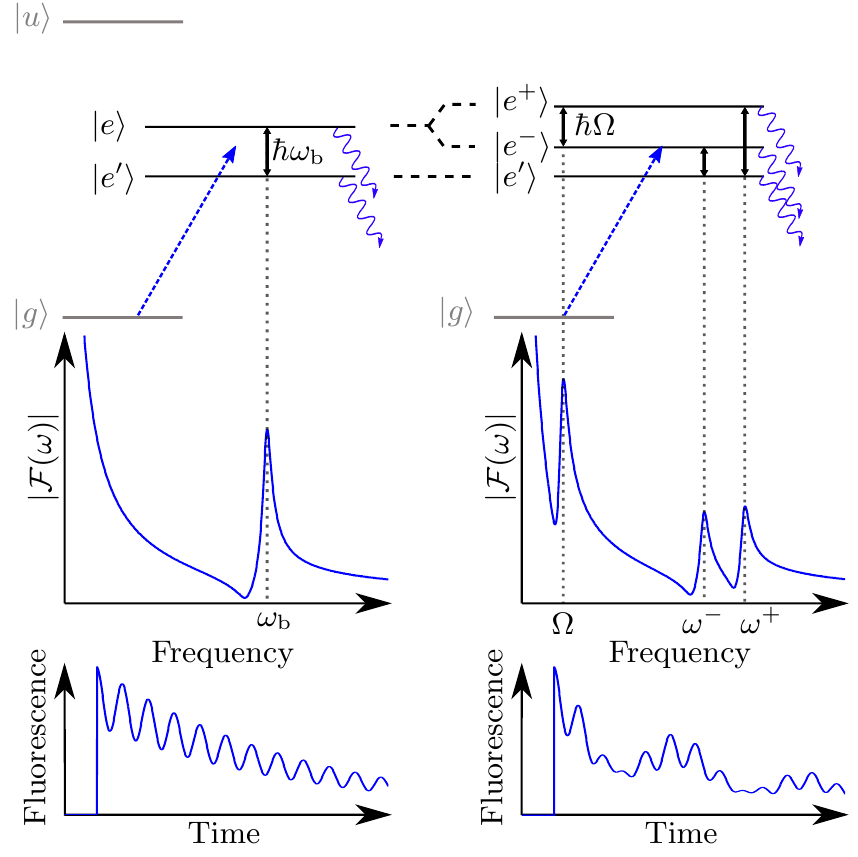}
\caption[]{\label{fig:toy_model}(Color Online)
Toy model of our experiment. Left: Atoms prepared in a superposition of closely spaced excited states $|e\rangle$ and $|e'\rangle$ demonstrate quantum beats at a frequency corresponding to the difference in their energies. Right: Driving an excited-state transition splits state $|e\rangle$ into two dressed states $|e^+\rangle$ and $|e^-\rangle$ and the dynamics of the excited-state transition are written into the quantum beats. We show the level scheme (top), the Fourier spectrum, $|\mathcal{F}(\omega)|$, of the fluorescence (middle) and the time-dependent fluorescence into an appropriately chosen polarization mode (bottom). The Fourier spectra are calculated by taking the magnitude of the Fourier transform of the  fluorescence signals.
}
\end{figure}

In this paper we make novel use of hyperfine quantum beats~\cite{haro73,hack91} to probe the excited-state transition dynamics of an `inverted-wavelength' ladder system in a thermal vapor. We find strong evidence for both Rabi oscillations and sub-Doppler Autler-Townes splitting. 

The paper is organized as follows:
In Section~\ref{sec:toy_model} we construct a toy model of our experiment, giving an overview of the physics involved. Section~\ref{sec:experiment} details our experimental procedure and in Section~\ref{sec:results} we present results in both the time-domain and the frequency domain. Section~\ref{sec:theory} outlines a computer model that we developed to understand the signals, which we  compare to the data in Section~\ref{sec:analysis}. The model yields good qualitative agreement which allows us to interpret features that we observe in the frequency domain.


\section{Principle of Perturbed Quantum Beats in a Ladder System}
\label{sec:toy_model}


\noindent In this section we outline a toy model of our ladder system which includes the minimum possible complexity to illustrate the physical principle (Figure~\ref{fig:toy_model}). The toy model considers a zero-velocity atom with ground state $|g\rangle$, an intermediate excited state $|e\rangle$ and an upper excited state $|u\rangle$. There is also a reference state $|e'\rangle$ which is close in energy to $|e\rangle$. The transition from $|g\rangle~\to~|e\rangle$ is driven by a short pulse whilst a continuous wave (CW) laser drives the excited-state transition from $|e\rangle~\to~|u\rangle$. For a sufficiently short excitation pulse the bandwidth exceeds the energy interval between $|e\rangle$ and $|e'\rangle$, and a coherent superposition of the two states is prepared by the pulse. The dynamics of the excited-state transition are read out by measuring the fluorescence from states $|e\rangle$ and $|e'\rangle$.

We begin our explanation by considering the simple case when the excited-state transition driving field is switched off (left column of Figure~\ref{fig:toy_model}). Once the coherent superposition of states $|e\rangle$ and $|e'\rangle$ has been prepared, the total fluorescence decays exponentially according to the state lifetime. However, the fluorescence into an appropriately chosen mode, characterized by polarization and propagation direction, is modulated by beating~\cite{haro76}. These `quantum beats' represent interference between the two different quantum pathways associated with $|e\rangle$ and $|e'\rangle$. The interference is erased if information regarding which pathway was taken is recovered (e.g. spectroscopically resolving the fluorescence from each state). In our toy model the time-dependent fluorescence into a particular mode has the form of an exponentially decaying envelope modulated by beating. The modulus of the Fourier transform of this time-dependent fluorescence, $|\mathcal{F}( \omega)|$, allows us to read off the beat frequency (see middle row of Figure~\ref{fig:toy_model}). States $|e\rangle$ and $|e'\rangle$ have energy $\hbar \omega_{e}$ and $\hbar \omega_{e'}$ respectively, and the beat frequency $\omega_{{\rm b}} = \omega_{e} - \omega_{e'}$ corresponds to the difference in energy. The visibility of the beats from a zero-velocity atom is set by a number of factors including the distribution of the population between the two states and the relative strengths with which the states couple to the selected fluorescence mode. In our experiment the visibility  is limited by velocity averaging as well.


To understand the effects of driving the excited-state transition it is easiest to consider the dressed state picture (right hand column of Figure~\ref{fig:toy_model}). CW driving of the excited-state transition splits $|e\rangle$ into two dressed states, $|e^+\rangle$ and $|e^-\rangle$, separated according to the Rabi frequency of the driving field, $\Omega$. The original beat at frequency $\omega_{{\rm b}}$, is split into two distinct beats with frequencies, $\omega^+~=~\omega_{{\rm b}}~+~\Omega/ 2$ and $\omega^-~=~\omega_{{\rm b}}~-~\Omega / 2$. Furthermore, a new beat frequency is introduced with frequency $\Omega$. This beat frequency relates to Rabi oscillations with atoms cycling on the excited-state transition. We note that unlike the initial quantum beat, this cycling leads to a modulation of the total fluorescence, not just a particular polarization mode. The Fourier spectrum, $|\mathcal{F}( \omega)|$, includes  all information regarding Autler-Townes splitting of the state $|e\rangle$ and Rabi oscillations on the excited-state transition, $|e\rangle~\to~|u\rangle$. The more complicated form of the time-dependent fluorescence is shown in the lower right panel of Figure~\ref{fig:toy_model}.

\begin{figure}
\includegraphics[width=3.4in]{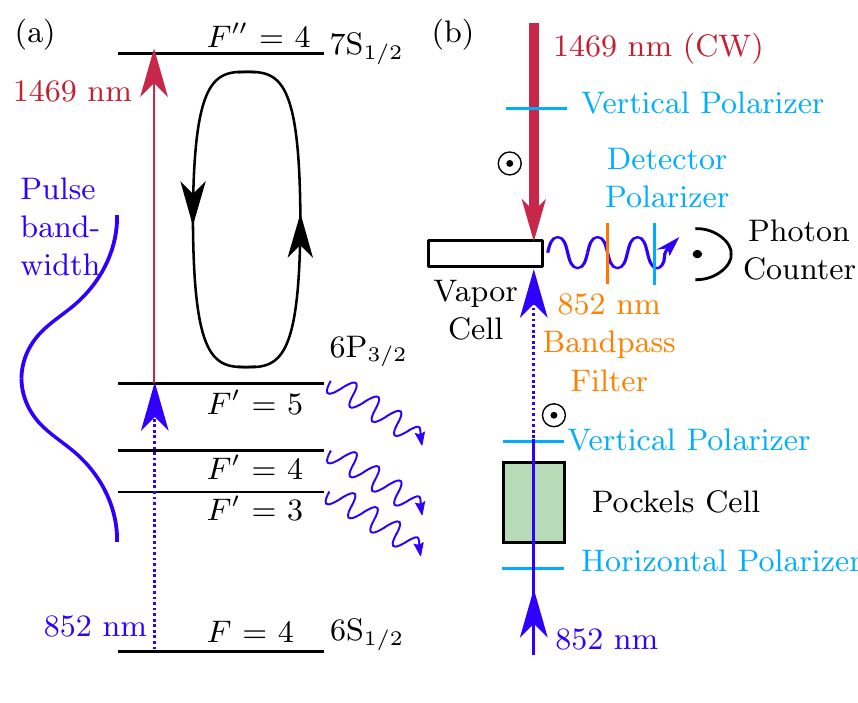}
\caption[]{\label{fig:Exp}(Color Online)
(a) Level scheme of our experiment: A short pulse of light excites several states in the 6P$_{3/2}$ manifold and a CW laser drives an excited-state transition 6P$_{3/2}$\,$F$\,=\,5~$\to$~7S$_{1/2}$\,$F$\,=\,4 (b) Schematic of experiment: Vertically polarized beams counter-propagate through a cesium vapor cell and fluorescence from the D$_2$ transition is detected with a single photon counter.
}
\end{figure}


\section{Experiment}
\label{sec:experiment}

\begin{figure*}
\includegraphics[width=\textwidth]{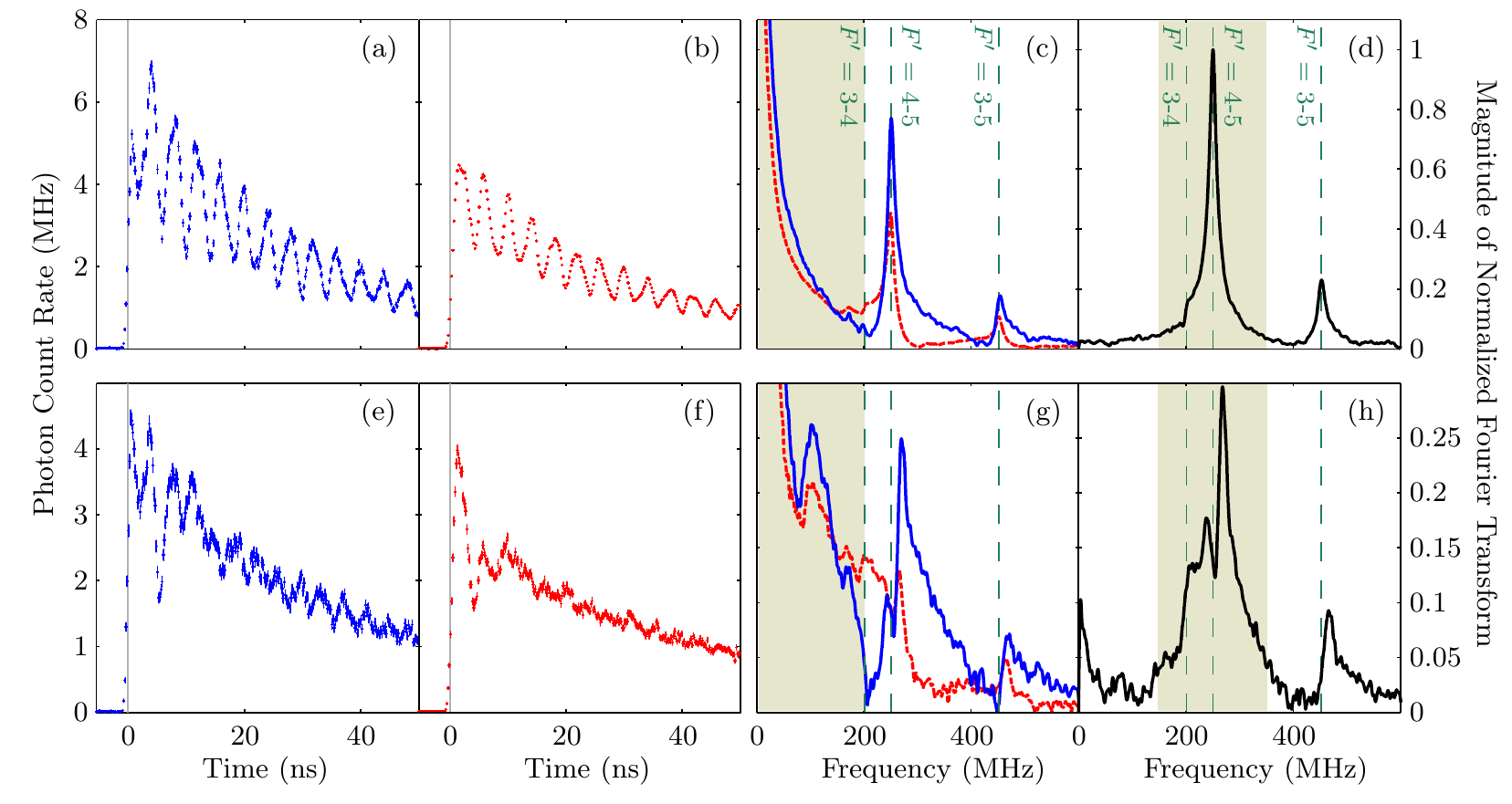}
\caption[]{\label{fig:TwoCol}(Color Online)
Top row (a-d): Measurements of fluorescence showing unperturbed hyperfine quantum beats. Bottom row (e-h): Measurements of fluorescence showing quantum beats that are modified by CW driving field with intensity $I_{\rm d}~=~4~$W\,cm$^{-2}$. We present measurements of time-dependent vertically polarized fluorescence~(a,e) and horizontally polarized fluorescence~(b,f). In panels (c,g) we present the Fourier spectra, calculated by taking the magnitude of the Fourier transform of the time-dependent fluorescence signals. The solid line (blue) shows the vertically polarized fluorescence and the dashed line (red) shows the horizontally polarized fluorescence. In panels (d,h) we show the spectrum of the difference between the two polarization signals. The spectra are normalized such that the peak in the difference signal relating to the unperturbed $F'\,=\,5\,\to\,4$ beat (d) has a height of~1 (See the Appendix for full details). The dashed vertical lines correspond to the 6P$_{3/2}$ hyperfine splitting~\cite{das05} and the shaded bands correspond to regions presented as color plots in Figure~\ref{fig:RawData}.
}
\end{figure*}

\noindent The simplified level scheme and experimental setup are shown in Figure~\ref{fig:Exp}a and Figure~\ref{fig:Exp}b respectively. We use cesium atoms in a vapor cell (length 2~mm) at room temperature (19\,$^{\circ}$C). The ladder scheme comprises the 6S$_{1/2}$\,$F$\,=\,4 state as the ground state, 6P$_{3/2}$\,$F'$~=\,5 as the intermediate state and 7S$_{1/2}$\,$F''$~=\,4 as the upper excited state. The other 6P$_{3/2}$ hyperfine states play the role of the reference state described in Section~\ref{sec:toy_model}.

We excite  the first transition using a short pulse (FWHM of 1~ns) of 852~nm  light generated by a CW diode laser stabilized to the $F\,=\,4\,\to\,F'\,=\,5$ hyperfine transition and modulated  by a Pockels cell between two crossed, high extinction polarizers. The short pulse duration means that the pulse bandwidth spans the hyperfine energy splitting of the 6P$_{3/2}$ manifold and therefore prepares a coherent superposition of several hyperfine states. It is this coherent superposition of states that leads to quantum beats in our system. The excited-state transition is driven by a counter-propagating, CW laser beam locked to the 6P$_{3/2}\ F'\,=\,5\,\to 7$S$_{1/2}\,F''\,=\,4$ (1469 nm) transition using excited-state polarization spectroscopy~\cite{carr12}. 

To best control the effects of driving the excited-state transition, it is desirable to minimize the spread of intensity of the excited-state transition driving field that the atoms experience. To achieve this, we only sample the center of the CW driving laser beam ($1/{\rm e}^2$ radius 0.3~mm) where the intensity is most uniform, by virtue of tighter focusing of the preparation pulse ($1/{\rm e}^2$ radius 0.06~mm). Both the laser beams are vertically polarized, and we detect fluorescence propagating in the horizontal plane. A narrow-band filter is used to select only fluorescence from the D$_2$ transition and a polarizer selects a particular mode of this fluorescence. 

The fluorescence is measured using a single photon detector module which generates a TTL level pulse for each photon. The pulses are timed and counted by a high-bandwidth oscilloscope and in this way we achieve nanosecond timing resolution. To avoid saturating the counting module, we ensure that the expected delay between photons is much longer than the dead time of the counting module ($\approx\, $35~ns).


\section{Results}
\label{sec:results}

\noindent We begin by considering the case of unperturbed hyperfine quantum beats. Figure~\ref{fig:TwoCol}(a) and Figure~\ref{fig:TwoCol}(b) show measurements of vertically and horizontally polarized fluorescence respectively, with the center of the excitation pulse incident at time $t=0$\,ns. As we noted in Section~\ref{sec:toy_model}, the total fluorescence is not modulated and so we see that the beating of the vertically and horizontally polarized fluorescence is out of phase. In Figure~\ref{fig:TwoCol}(c) we present the magnitudes of the normalized Fourier transforms of these fluorescence measurements and in Figure~\ref{fig:TwoCol}(d) we remove frequency components relating to the exponential decay envelope by subtracting the two signals (See the Appendix for details of normalization and subtraction). Because the beating of the two polarization signals is out of phase we retain the quantum beat frequency components and so we observe peaks at 201,~251~and~452~MHz, corresponding to the 6P$_{3/2}$ hyperfine splitting~\cite{das05} (highlighted with vertical dashed lines). The peak relating to the $F'=3\to F'=4$ quantum beat (201~MHz) is very weak as the population in these two states is limited. This restricted population is a result of both weaker coupling to the ground state and also detuning from the middle of the excitation pulse bandwidth which is centered on the $F=4\to F'=5$ transition.

When we drive the excited-state transition the quantum beats are modified. For the driving field intensity at the center of the laser beam $I_{\rm d}\,=\,4\,$~W\,cm$^{-2}$, we present the vertically and horizontally polarized fluorescence measurements in Figure~\ref{fig:TwoCol}(e) and Figure~\ref{fig:TwoCol}(f) respectively, along with their Fourier spectra in Figure~\ref{fig:TwoCol}(g) and the spectrum of the difference signal in Figure~\ref{fig:TwoCol}(h). We can see the changes to the Fourier spectra that we expected from considering the toy model in Section~\ref{sec:toy_model}. Firstly the peak relating to the $F'=5\to F'=4$ beat (251~MHz) is split in two (Figure~\ref{fig:TwoCol}(g,h)). The origin of this effect is Autler-Townes splitting of the 6P$_{3/2}~F'=5$ atomic state, caused by driving the excited-state transition. Secondly a new oscillation is present, leading to a peak at 100~MHz in this example (Figure~\ref{fig:TwoCol}(g)). This represents atoms performing Rabi oscillations on the excited-state transition. The absence of this peak from the difference signal in Figure~\ref{fig:TwoCol}(h) is because the Rabi oscillations modulate the entire 852~nm fluorescence. Therefore the oscillation is in phase between the vertically and horizontally polarized fluorescence and is removed in the difference signal.

It is interesting to note that whilst the simple model outlined in Section~\ref{sec:toy_model} predicts that the splitting of the beat frequency would be equal to the frequency of the Rabi oscillation, it is clear from Figure~\ref{fig:TwoCol}(g) that this is not the case. The cause of this discrepancy stems from Doppler effects which we explore and explain in Section~\ref{sec:analysis} using a comprehensive computer simulation outlined in Section~\ref{sec:theory}. 

\begin{figure}
\includegraphics[width=3.4in]{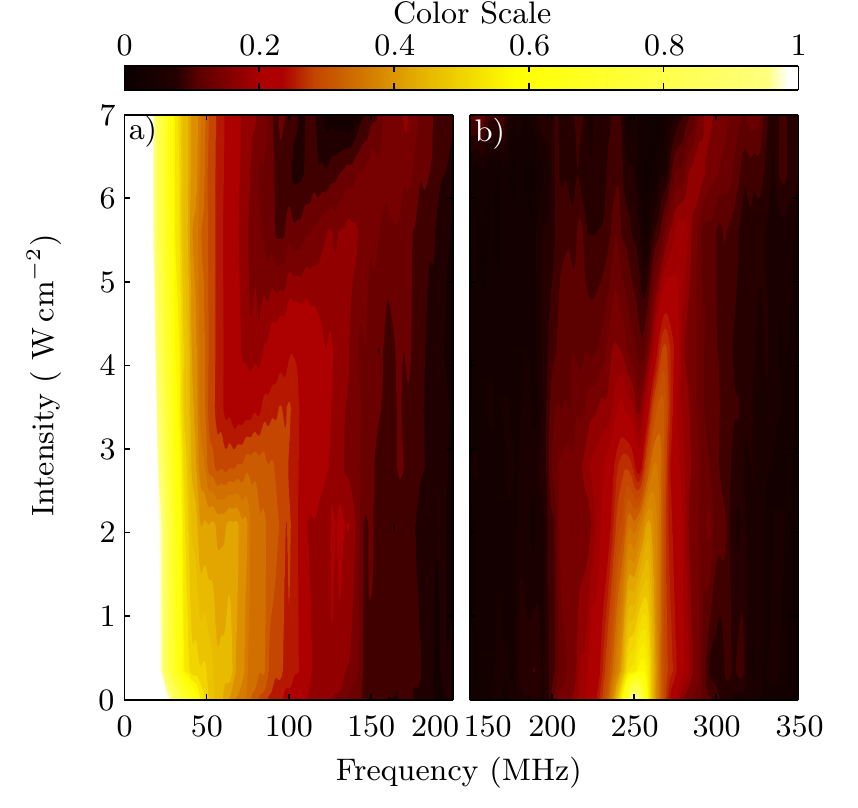}
\caption[]{\label{fig:RawData}(Color Online) 
Colorplots of the magnitude of the Fourier transforms of the fluorescence measurements. (a)~Vertically polarized fluorescence shows a diagonal feature that corresponds to Rabi oscillations. (b)~The difference signal demonstrates a branched feature relating to Autler-Townes splitting. Parts (a) and (b) relate to the highlighted regions of Figure~\ref{fig:TwoCol}(c,g) and Figure~\ref{fig:TwoCol}(d,h) respectively.
}
\end{figure}

In Figure~\ref{fig:RawData} we present colorplots covering a range of excited-state transition laser driving intensities $I_{\rm d}\,=\,0\to\,7\,$~W\,cm$^{-2}$, constructed from nine individual sets of intensity measurements. Figure~\ref{fig:RawData}(a) shows the modulus of the Fourier transform of the vertically polarized fluorescence measurements. The diagonal feature corresponds to the Rabi oscillation, which increases in frequency with increasing laser power. Figure~\ref{fig:RawData}(b) shows the modulus of the Fourier transform of the difference signal and the splitting of the $F'=5\to F'=4$ beat into two separate branches is clear.

We also point out some further, more subtle effects. First, the fluorescence decays more slowly as the longer lived 7S$_{1/2}~F''\,=\,4$ state is mixed into the 6P$_{3/2}$ states. Second, the total amount of measured 852~nm fluorescence decreases. This is partly because the atoms can now decay from the 7S$_{1/2}~F''\,=\,4$ state via the 6P$_{1/2}$ manifold as well as the 6P$_{3/2}$ manifold, but could also be due to hyperfine optical pumping caused by light leaking through the Pockels cell between pulses. Further consequences of this effect are discussed in Section~\ref{sec:analysis}. Finally we note that the higher frequency branch of the split $F'=5\to F'=4$ quantum beats is stronger than the low frequency branch (Figure~\ref{fig:RawData}(b)). This effect is even more exaggerated in the $F'=5\to F'=3$ (452 MHz) beat where we do not observe the low frequency branch at all (Figure~\ref{fig:TwoCol}(d,g)). The absence of the lower branch originates from Doppler effects that we also discuss in Section~\ref{sec:analysis}.


\section{Computer Model}
\label{sec:theory}

\noindent Here we develop a theoretical simulation to predict the behavior of our system. Conceptually, it involves two steps. First the optical Bloch equations for the system are solved numerically; second, the time-dependent expectation value of a  `detection operator', $\mathcal{B}$, is calculated, giving the expected fluorescence \cite{haro76}. The operator has the form,

\begin{equation}
\label{eq:det}
\mathcal{B} = C \sum_{f} \textbf{e. \^{D}}  |f\rangle \langle f|\textbf{e*. \^{D}},
\end{equation}

\noindent where $C$ is a coefficient relating to detection efficiency, \textbf{e} is a unit vector describing the polarization of the detected fluorescence and \textbf{\^{D}} is the electric dipole operator for the D$_2$ transition of the atom. The final states $f$ include all of the magnetic sub-levels ($m_{{\rm F}}$) of the two 6S$_{1/2}$ hyperfine ground states. We note that the expectation value of the operator $\mathcal{B}$ is proportional to the square of the atomic dipole projected onto the detected polarization angle and measures the coupling of the atomic state to the field modes. The calculation process is repeated for a sample of velocity classes which are then summed and weighted according to a Boltzmann distribution.

The computation basis is fixed such that the linearly polarized excitation lasers drive only $\pi$ transitions, allowing our calculation to be performed in a set of mutually uncoupled $m_{{\rm F}}$ subspaces. Note that this basis might not be the energy eigenbasis due to uncompensated laboratory magnetic fields. However, any coherence developed between the $m_{{\rm F}}$ subspaces as a consequence of this can be neglected since the duration of our experiment is much shorter than the relevant Larmor precession timescale. The time evolution of the density matrix $\hat{\rho}_{m_{{\rm F}}}$ in each of the nine $m_{{\rm F}}$ subspaces, is calculated using a set of optical Bloch equations,

\begin{equation}
\label{eq:Lou}
\dot{\hat{\rho}}_{m_{\rm F}} = \frac{i}{\hbar}\begin{bmatrix}\hat{\rho}_{m_{\rm F}},\hat{H}_{m_{\rm F}}\end{bmatrix} - \hat{\Gamma},
\end{equation}

\noindent where $\hat{H}_{m_{\rm F}}$ is the Hamiltonian for each subspace, and $\hat{\Gamma}$ is a decay operator. The Rabi frequencies are calculated individually for each $m_{\rm F}$ subspace and each subspace includes one state from each of the hyperfine levels: 6S$_{1/2}~F =~3,4$; 6P$_{3/2}~F'~=~3,4,5$ and 7S$_{1/2}~F''~=~4$. This convenient sub-division offers a computational speed up that permits the simulation to be run on a desktop computer.

Although the subspaces are not coupled by the driving laser fields, we note that they are not truly separate, since atoms can undergo spontaneous $\sigma^{\pm}$ transitions resulting in a change of $m_{\rm F}$ quantum number. Instead of modeling this full behavior, we attribute the total rate of spontaneous decay of each state to $\pi$ transitions, thus conserving the total population in each subspace. In this way, we are able to capture the lifetimes of the states and retain the computational efficiency. Furthermore, because the timescale of our experiment is set by a single atomic state lifetime, we are confident that the effect of these angular momentum changing processes is negligible, as there is insufficient time to redistribute atomic population amongst the $m_{\rm F}$ subspace.

In the final step of our model, we collate the populations and coherences from the nine subspaces into a single density matrix, giving the complete state of the atom as it changes in time. Using the `detection operator', we project the atomic dipole at each time step and hence infer both the linear and circularly polarized 852~nm fluorescence. 

\begin{figure}
\includegraphics[width=3.4in]{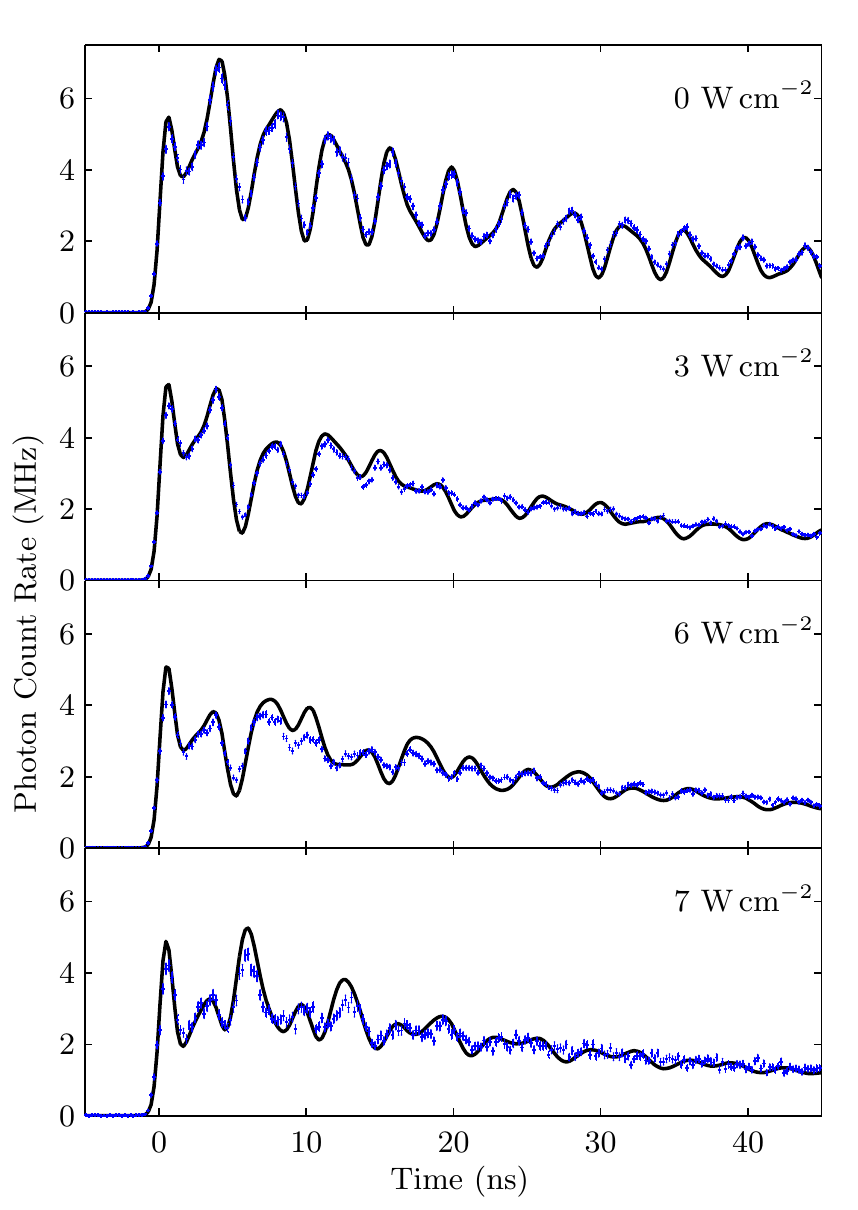}
\caption[]{\label{fig:Comp}(Color Online)
Vertically polarized fluorescence: We compare the model (black line) and experimental data points (blue) for measured excited-state transition laser intensity $I_{{\rm d}}$\,=\,(0,\,3,\,6,\,7)~${\rm W\,cm}^{-2}$ (top to bottom). We note that the visibility of the peaks is always smaller than the model predicts. The incident light pulse occurs at time $t$~=~0~ns and the error bars are calculated from Poissonian photon counting statistics~\cite{hughes}.
}
\end{figure}


\section{Analysis}
\label{sec:analysis}

\noindent In this section we make a direct comparison between the computer simulation and the measured data. In Figure~\ref{fig:Comp} we present the results of the vertically polarized fluorescence for both the experiment and simulation. The unperturbed hyperfine quantum beat signal fits well, and we see at least qualitative agreement for the perturbed beats. Although the features are often more pronounced in the simulation than the data, there is a qualitative match between the data and theory. On the strength of this we can draw additional physical insight about the system.

\begin{figure}
\includegraphics[width=3.4in]{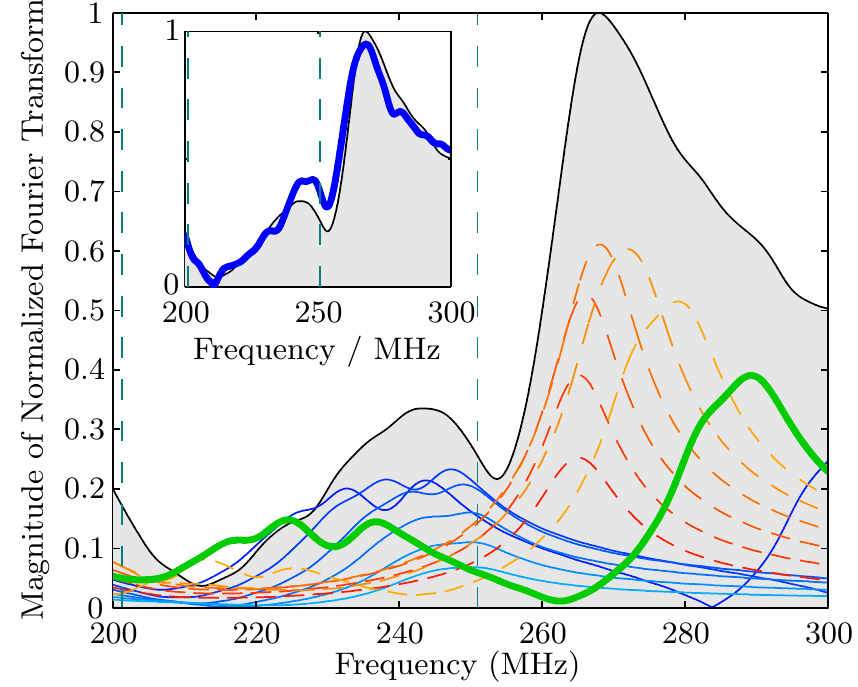}
\caption[]{\label{fig:FillIn}(Color Online)
Calculated magnitude of the normalized Fourier transform of the vertically polarized fluorescence: We show a break down of velocity class contributions, with dashed (red) lines showing individual velocity classes traveling away from the 852~nm laser, and solid (blue) lines showing velocity classes traveling towards the 852~nm laser. The velocity classes are spaced at 33~m\,s$^{-1}$ intervals and the bold (green) line shows the contribution from zero-velocity atoms.  The gray shaded area shows the scaled sum of the signals and the calculation relates to probing a region with a uniform excited-state transition driving field intensity $I_{\rm d}^{{\rm Sim}}\,=\,0.9\,$\,W\,cm$^{-2}$. The inset compares the data taken for measured CW driving field intensity $I_{\rm d}\,=\,3\,$W\,cm$^{-2}$ shown in bold (blue) and the summed model (gray shaded area and black line).}
\end{figure}

In Section~\ref{sec:results} we noted that the splitting of the $F'\,=\,5\to\,F'\,=\,4$ hyperfine quantum beat was unexpectedly smaller than the measured frequency of the Rabi oscillation. We suggest this is similar to narrowed EIT windows in thermal vapors \cite{Natarajan08,baso09} where off-resonant velocity classes partially fill the transparency window left by resonant atoms. Calculated contributions to the splitting of the $F'~=~4~\to~F'~=~5$ quantum beat from different velocity classes are shown in Figure~\ref{fig:FillIn}. The zero-velocity class (bold, green) shows a splitting that is consistent with the simulated excited-state transition Rabi frequency, yet this is much larger than the splitting which appears in the total signal. Figure~\ref{fig:FillIn} shows how contributions from off-resonant velocity classes fill in the gap. The inset compares this best fit calculated spectrum with our data and we see qualitative agreement, although we acknowledge a significant discrepancy in the simulated ($I_{\rm d}^{{\rm Sim}}\,=\,0.9$~W\,cm$^{-2}$) and measured ($I_{\rm d}\,=\,3$\,W\,cm$^{-2}$) excited-state transition laser intensities.

We also noted in Section~\ref{sec:results} that the high-frequency branch of the split $F'~=~4~\to~F'~=~5$ quantum beat makes a stronger contribution to the Fourier spectrum than the low-frequency branch. This unexpected asymmetry can be explained by constructing a two-step argument: First, we note that the strongest beats arise from atoms experiencing a red Doppler shift of the 852~nm laser. This moves the center of the frequency profile of the pulse between the two beating transitions, promoting the excitation of both levels as required  for quantum beats. Second, the atoms which experience a red-shift for the 852~nm laser see a blue-shift of the 1469~nm laser because the laser beams are counter-propagating. This blue-shift means that the dressed states represented in the higher frequency branch of the split quantum beat have a greater admixture of the 6P$_{3/2}~F'$\,=\,5 state, and as such a stronger coupling to the ground state. Thus the imbalance between branches of the quantum beat comes from a bias towards a particular velocity class, and a bias within this velocity class to a particular branch. The asymmetry between branches is even stronger for the $F'~=~3~\to~F'~=~5$ quantum beat as the $F'\,=\,3$ hyperfine state is further in energy from the $F'\,=\,5$ hyperfine state. Consequently, only the high frequency branch of the splitting was observed and the lower-frequency branch is absent. (Section~\ref{sec:results}).

There are some remaining discrepancies between the simulation and the data. We found that when we used the CW driving field intensity as a fit parameter in the model, the best fit did not match or even scale linearly with the intensity we measured in the experiment. We believe that this might originate from optical pumping between excitation pulses. The high extinction polarizers each side of the Pockels cell (Figure~\ref{fig:Exp}b) still allowed a few hundred nanowatts of 852~nm light to leak into the vapor cell for the 1~ms duration between pulses. For resonant velocity classes, this could have lead to an initial state other than the uniform distribution over the ground states that the computer simulation assumes.


\section{Conclusion}
\label{sec:conclusion}

\noindent We have demonstrated a novel method using hyperfine quantum beat spectroscopy for observing sub-Doppler Autler-Townes type splitting in an `inverted wavelength' ladder scheme which would not be observable in a continuously excited room-temperature vapor. A comprehensive model of the fluorescence gives qualitative agreement with our data, and we use it to gain physical insight into the process. By exploiting our method to its full potential it would be possible to combine information from both the Autler-Townes splitting and the Rabi oscillations to achieve a complete read out of excited-state transition dynamics. Our work on ladder excitation schemes contributes to a general effort towards the exploitation of Rydberg atoms in a room temperature atomic vapor using multi-photon, step-wise excitation. In a wider context, our method offers a new means for investigating excited-state transitions in a room-temperature vapor.


\begin{acknowledgements}
The authors would like to thank Ifan Hughes for stimulating discussions and acknowledge financial support from EPSRC [grant EP/K502832/1] and Durham University. The data in this paper are provided in the Supplemental Material
\end{acknowledgements}

\appendix
\section{Normalization of Fourier Spectra}

\noindent In Figure~\ref{fig:TwoCol} we present normalized Fourier spectra of the time domain signals. Here we inform the reader of the details of the normalization. 

We begin with the measured time domain quantities. For unperturbed beats, $h_0(t)$ and $v_0(t)$ correspond to the horizontally and vertically polarised fluorescence respectively, and for the modified quantum beats $h_m(t)$ and $v_m(t)$ correspond to the horizontally and vertically polarised fluorescence respectively. These quantities are Fourier transformed to give $H_0(\omega)$, $V_0(\omega)$, $H_m(\omega)$ and $V_m(\omega)$. The horizontal and vertical signals are scaled and subtracted to give the difference signals, $D_0(\omega)$ and $D_m(\omega)$.



\begin{equation}
\label{eq:D1}
D_0(\omega) = H_0(\omega) - \frac{\sum{h_0(t)}}{\sum{v_0(t)}} V_0(\omega),
\end{equation}
 
\noindent and

\begin{equation}
\label{eq:D0}
D_m(\omega) = H_m(\omega) - \frac{\sum{h_m(t)}}{\sum{v_m(t)}} V_m(\omega),
\end{equation}

\hspace{5cm}

\noindent where $D_0(\omega)$ and $D_m(\omega)$ correspond to unperturbed beats and modified beats respectively. Scaling the vertically polarised fluorescence signal in this way means that the average of the difference signals is zero in the time domain.

A global normalization factor, $n$, is found using $D_0$, corresponding to the height of the $F'\,=\,4\,\to\,5$ (251~MHz) unperturbed quantum beat.

\begin{equation}
\label{eq:n}
n = |D_0(\omega_{F'\,=\,4\,\to\,5})|,
\end{equation}

\noindent Finally we plot the quantities $H_0'(\omega)$, $V_0'(\omega)$, $D_0'(\omega)$, $H_m'(\omega)$, $V_m'(\omega)$ and $D_m'(\omega)$, corresponding to the normalized modulus of the Fourier spectra, such that

\begin{equation}
\begin{array}{r@{}l}
H_0'(\omega) \ &= \ |H_0(\omega)| / n,
\\
V_0'(\omega)  \ &= \ |V_0(\omega)| / n,
\\
D_0'(\omega)  \ &= \ |D_0(\omega)| / n,
\\
H_m'(\omega)  \ &= \ |H_m(\omega)| / n,
\\
V_m'(\omega)  \ &= \ |V_m(\omega)| / n,
\\
D_m'(\omega)  \ &= \ |D_m(\omega)| / n.
\end{array}
\end{equation}

As a result of this normalization, the peak in the difference signal relating to the unperturbed $F'\,=\,5\,\to\,4$ beat has a height of~1, and the results from different excited-state transition driving field strengths are directly comparable.


\end{document}